\documentclass[runningheads]{llncs}
\usepackage[T1]{fontenc}
\usepackage{graphicx,verbatim}
\usepackage[T1]{fontenc}
\usepackage{graphicx}
\usepackage{diagbox}
\usepackage{amsfonts}
\usepackage{multirow}
\usepackage{enumitem}
\usepackage{amsmath}

\usepackage{hyperref}
\usepackage[capitalise]{cleveref}
\usepackage{xcolor,soul}
\hypersetup{
    colorlinks=true,
    linkcolor=blue,
    filecolor=cyan,      
    urlcolor=magenta,
    }
\usepackage{booktabs}
\usepackage{color}

\usepackage{bbm}
\usepackage{float}
\usepackage{color, colortbl}
\usepackage{amssymb}  
\usepackage{pifont}   
\usepackage{subcaption} 
\usepackage{tikz}  
\usepackage[inkscapelatex=false]{svg}
\begin{document}
\title{Modality-Agnostic Brain Lesion Segmentation with Privacy-aware Continual Learning}
%
\author{Yousef Sadegheih \inst{1} \and
Pratibha Kumari\inst{1} \and
Dorit Merhof\inst{1,2}}


%
\authorrunning{Y. Sadegheih et al.}
%
\institute{Faculty of Informatics and Data Science, University of Regensburg, Regensburg, 93053, Germany \and
Fraunhofer Institute for Digital Medicine MEVIS, Bremen 28359, Germany
\email{dorit.merhof@ur.de}\\
}




\maketitle              
\begin{abstract}
Traditional brain lesion segmentation models for multi-modal MRI are typically tailored to specific pathologies, relying on datasets with predefined modalities. Adapting to new MRI modalities or pathologies often requires training separate models, which contrasts with how medical professionals incrementally expand their expertise by learning from diverse datasets over time. Inspired by this human learning process, we propose a unified segmentation model capable of sequentially learning from multiple datasets with varying modalities and pathologies. Our approach leverages a privacy-aware continual learning framework that integrates a mixture-of-experts mechanism and dual knowledge distillation to mitigate catastrophic forgetting while not compromising performance on newly encountered datasets. Extensive experiments across five diverse brain MRI datasets and four dataset sequences demonstrate the effectiveness of our framework in maintaining a single adaptable model, capable of handling varying hospital protocols, imaging modalities, and disease types. Compared to widely used privacy-aware continual learning methods such as LwF, SI, EWC, MiB, and TED, our method achieves an average Dice score improvement of approximately 14\%. Our framework represents a significant step toward more versatile and practical brain lesion segmentation models, with implementation available on \href{https://github.com/xmindflow/BrainCL}{github.com/xmindflow/BrainCL}.

\keywords{Continual learning  \and Variable MRI modality \and Brain lesion.}

\end{abstract}

\section{Introduction}\label{sec:intro}
\begin{figure*}[!t]
\centering
\includegraphics[width=\linewidth]{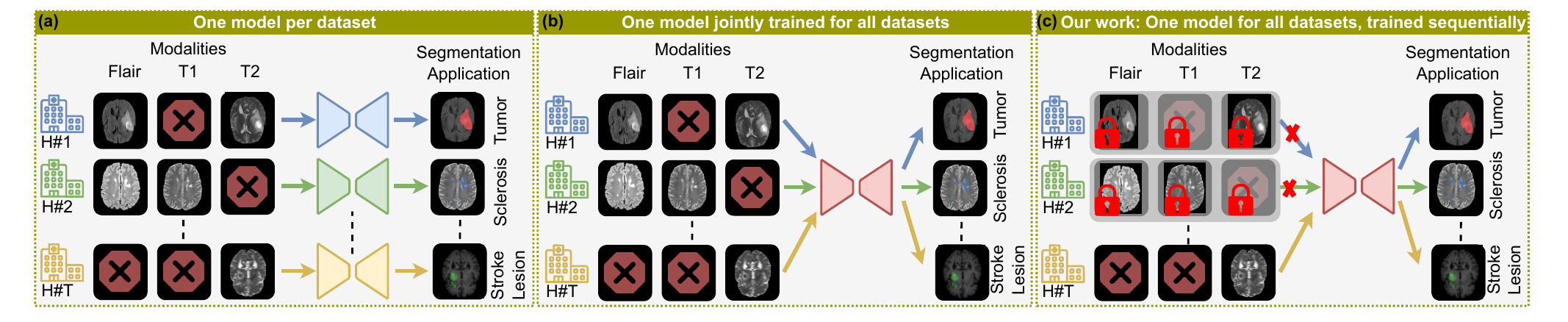}
\caption{Training paradigms: (a) separate models for fixed modality sets, (b) joint training, and (c) our CL framework for sequential training without past data.}
    \label{fig:CL_yousef_proposed}
\end{figure*}

Magnetic Resonance Imaging (MRI)-based brain lesion segmentation is crucial in neurology for analysis, surgery planning, and functional imaging. However, real-world clinical applications face challenges due to patient, scanner, and pathology variability. Traditionally, UNet-based models are trained for specific pathologies with fixed modalities, limiting flexibility. This often requires training separate models for different modality-pathology combinations (Fig. \ref{fig:CL_yousef_proposed}a), which is resource-intensive and less flexible. In contrast, clinicians adapt to different diseases and modalities. Likewise, a single model learning from diverse datasets can enhance performance by leveraging pathology relationships, especially for small datasets.
Recent studies \cite{xu2024feasibility,wagner2024feasibility} train a single UNet on multiple datasets with variable modalities (Fig. \ref{fig:CL_yousef_proposed}b). While this enables multi-dataset segmentation, it requires all datasets to be available simultaneously. Moreover, performance drops if test data differ in hospital, disease type, or lesion size.

This work addresses variable modality MRI segmentation, where datasets arrive sequentially rather than all at once (Fig. \ref{fig:CL_yousef_proposed}c). Continual Learning (CL) enables a single model to learn new datasets while retaining past knowledge \cite{mccloskey1989catastrophic}. Naively updating a UNet-based model disrupts previous weights, causing Catastrophic Forgetting (CF). CL prevents this by using strategies such as storing past data, penalizing larger weight changes, or allocating parameter subsets datasets. CL is gaining interest in medical image analysis \cite{Kum_Continual_MICCAI2024,kumari2023continual} and, more specifically, also in studies for brain MRI segmentation under domain shifts \cite{karani2018lifelong,van2019towards}. 
However, modality variability remains unexplored.
We improve upon \cite{xu2024feasibility} by enabling CL in 3D-UNet without requiring all datasets to be available simultaneously. Our buffer-free CL approach learns from diverse brain MRI datasets from different hospitals and pathologies. It combines dual-distillation-based regularization with soft parameter isolation for domain adaptation. The dual-distillation method transfers knowledge from the previous model at the feature and response levels to the new model trained on incoming data. Additionally, we integrate mixture-of-experts (MoE) \cite{shazeer2017outrageously} within each encoder and decoder layer of UNet to minimize interference between datasets. These experts are activated differently for each data distribution using a domain token, a binary vector encoding modality, and pathology information. This targeted activation mechanism helps the model retain knowledge from past datasets more effectively.
We summarize our contributions as follows: 
\ding{182} To the best of our knowledge, this is the first study exploring CL for brain MRI segmentation under domain shifts, including heterogeneous modalities, pathologies, and acquisition centers.
\ding{183} We introduce a novel domain-conditioned MoE in UNet, incorporating modality and pathology information for 3D segmentation.
\ding{184} Our dual-distillation and MoE-based CL strategy outperforms existing buffer-free CL methods.

\section{Methodology}

\begin{figure*}[!t]
\centering
\includegraphics[scale=0.8]{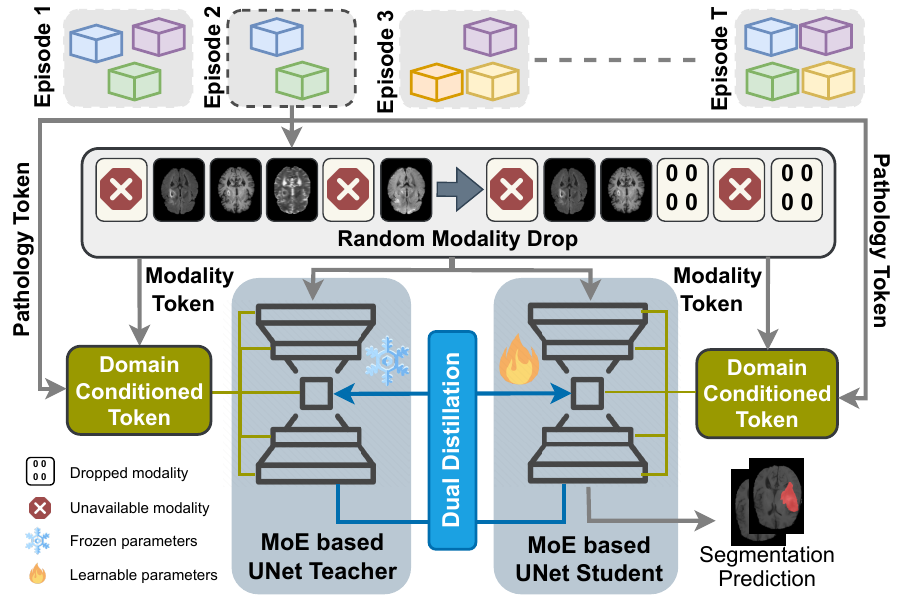}
\caption{Overview of proposed privacy-aware continual learning framework.
}
\label{fig:CL_yousef_framework}
\end{figure*}

In CL, a model sequentially learns datasets ($D_1, D_2, \dots D_T$), each differing in disease type, modalities, and data sources. At any time $t$, only the training set of $D_t$ is available, while test sets from all past datasets remain accessible. To prevent CF without storing past data, we use a buffer-free approach essential for privacy-sensitive applications. Our method employs dual knowledge distillation, where the previous model (teacher, $\mathcal{M}_{t-1}$) guides the current model (student, $\mathcal{M}_{t}$), ensuring knowledge preservation while learning new data. We also integrate a domain-conditioned MoE in convolution layers to minimize interference. The training combines a segmentation loss on the current dataset with dual distillation losses from the teacher model. Additionally, random modality dropout enhances generalization by exposing the model to different modality combinations. A flowchart of our approach is shown in Fig.~\ref{fig:CL_yousef_framework}. The next sections detail handling modality variations and our buffer-free CL strategy.
\subsection{Variable modality handling}

Segmentation datasets in a sequence ($D_1, D_2, \dots, D_T$) often have distinct MRI modality sets ($N_1, N_2, \dots, N_T$), where $N_t$ represents the modalities available in dataset $D_t$. Training with $D_t$ requires a UNet with $|{N_t}|$ input channels, necessitating a separate UNet for each unique modality set. However, for continual learning across datasets, a single model must handle all modality variations. 
A simple yet effective solution~\cite{xu2024feasibility} uses a single UNet with input channels set to the maximum number of expected modalities $K$, representing possible modalities as ${m_1, m_2, \dots, m_K}$. This design ensures compatibility with all datasets having $|{N_t}|\leq K$. If a modality $m_k$ is missing in a dataset, the corresponding channel is zero-filled. To improve generalization and mitigate spurious correlations between datasets and modalities, we employ random modality dropping during training. This exposes the model to varied modality combinations from each dataset, enhancing robustness. While this approach enables handling variable modalities within a unified framework, it is expected to slightly underperform compared to dedicated UNet models tailored for individual datasets.

\subsection{Dual distillation}
We propose a dual Knowledge Distillation (KD) strategy, transferring knowledge from the teacher to the student model at both latent features and response outputs. This dual alignment preserves structural and contextual information across hierarchies, enhancing the model's ability to retain learned knowledge. 
In the response-based KD~\cite{li2024continual}, student model is enforced to produce similar output (response) as the teacher model. Typically, KL-divergence between the teacher and student model outputs on current data is used as a regularization term, defined as:
\begin{equation}
\mathcal{L}^{\text{KLD}}= \text{KL} \left ( \sigma ({\mathcal{M}_{t}(x)}/ \tau )|| \sigma ({\mathcal{M}_{t-1} (x)}/ \tau )\right ) 
\label{eq:loss_kld}
\end{equation}
where $\text{KL}$ is KL-divergence, $\mathcal{M}_{t}(x)$ and $\mathcal{M}_{t-1}(x)$ are the output logits for input $x$ by $\mathcal{M}_{t}$ and $\mathcal{M}_{t-1}$, $\sigma$ is softmax operator applied to obtain soft targets for KD with temperature $\tau$. 
Instead of using a static regularization coefficient as in prior works~\cite{radio3,kirkpatrick2017overcoming}, we propose a dynamic coefficient ($\alpha_t$) for $\mathcal{L}^{\text{KLD}}$, that adapts based on the degree of domain shift between datasets. 
Greater shifts necessitate stronger regularization to mitigate forgetting. The shift is estimated using the inverse of the Dice Similarity Coefficient (DSC) on unseen data, scaled to a user-defined range $[\alpha_{\text{min}},\alpha_{\text{max}}]$ as $(\alpha_{min} + (1-\text{DSC}^{M_{t-1}}_{D_t})( \alpha_{\text{max}}-\alpha_{\text{min}}))$. This dynamic adjustment accounts for variability in dataset distributions, ensuring effective KD.

To complement response-based KD, we add latent-based KD~\cite{li2024continual}, which targets the alignment of latent representations between the student and teacher models. To achieve this, we employ a cosine similarity-based regularization loss, $\mathcal{L}^{\text{Cosine}}$, which aligns the latent feature representations of the student model with those of the teacher model

\begin{equation}
\mathcal{L}^{\text{Cosine}}= 1 - \frac{\mathbf{f}_{\mathcal{M}_{t-1}} \cdot \mathbf{f}_{\mathcal{M}_{t}}}{||\mathbf{f}_{\mathcal{M}_{t-1}}|| \, ||\mathbf{f}_{\mathcal{M}_{t}}||} ,
\label{eq:loss_cosine}
\end{equation}
where $\mathbf{f}_{\mathcal{M}_{t-1}}$ and $\mathbf{f}_{\mathcal{M}_{t}}$ represent flattened bottleneck features from the teacher model $\mathcal{M}_{t-1}$ and student model $\mathcal{M}_{t}$, respectively.

\subsection{Mixture-of-Expert}
When learning a new dataset naively, the model often overwrites previously acquired weights, leading to poor performance on old data. To address dataset conflicts, generic CL literature has mainly explored adding parameter subsets or reserving parameters in fixed networks~\cite{de2021continual}. However, parameter addition can cause unbounded model growth, while hard reservation restricts knowledge sharing across datasets. To overcome these issues, we draw inspiration from soft parameter reservation techniques in multi-task learning~\cite{shazeer2017outrageously}, which enable flexible capacity sharing across tasks. Specifically, we propose integrating a domain-conditioned MoE mechanism into each convolutional layer of the UNet model. This approach dynamically activates experts with varying strengths based on the domain, achieving soft parameter isolation in a fixed parameter network. The MoE uses a domain-conditioned token derived from dataset-specific metadata to compute gating weights via a linear gating network, $g=\sigma(W_g \textbf{c}+ b_g)$, where $\sigma$ is softmax operation, $W_g \in \mathbb{R}^{e \times (m+d)}$ and $b_g \in \mathbb{R}^{e}$ are the gating network's parameters for $e$ number of experts, and $\textbf{c} \in \mathbb{R}^{(m+d)}$ is a domain-conditioned token concatenating binary representations of available modalities $\textbf{I}^m$ and disease $\textbf{I}^d$, with $m$ and $d$ being maximum number of allowed modalities and pathology. This design ensures that each domain is handled uniquely, enabling the model to adaptively allocate expertise. Finally, the gating weights, $g=[g_1, g_2, \dots, g_e]$, are used to aggregate outputs from $e$ experts, $\{E_1, E_2, \dots, E_e\}$, as: $y=\sum_{i=1}^{e} g_i \cdot E_i(f)$, where $f$ is the input feature map, $E_i(\cdot)$ is the $i^{th}$ expert, and $y$ is the aggregated output. Unlike hard expert selection, our MoE employs soft selection, allowing contributions from all experts while dynamically adjusting their influence based on the dataset context.

\subsection{Model objective}
The student model is trained with joint supervision from the teacher model and current data, enabling it to perform effectively on both previously seen and newly introduced data.
At any session $t$, the student model $\mathcal{M}_t$ is optimized using a total loss $\mathcal{L}^{Total}$, comprising segmentation loss ($\mathcal{L}^{\text{Task}}$), cosine similarity loss ($\mathcal{L}^{Cosine}$), and KL-divergence loss ($\mathcal{L}^{KLD}$). The segmentation loss integrates Dice loss (Dice) and cross-entropy loss (CE), as commonly adopted in literature~\cite{sadegheih2024lhu}. This multi-component loss enables the model to balance between learning new information and retaining prior knowledge:

\begin{equation}
\mathcal{L}^{\text{Total}}= 
\underbrace{{\mathcal{L}^{\text{Dice}} +   \mathcal{L}^{CE} }}_{\mathcal{L}^{\text{Task}}}
+ \beta  \mathcal{L}^{\text{Cosine}} +\alpha_t  \mathcal{L}^{KLD}  .
\label{eq:loss_total}
\end{equation}

\section{Experimental setup and results}\label{sec:exp}

\subsection{Datasets, experimental setup and evaluation metrics} 
CL experiments are carried out on five brain MRI datasets (BRATS-Decathlon~\cite{bakas2017advancing}, MSSEG \cite{commowick2018objective}, ATLAS v2.0 \cite{liew2022large}, WMH \cite{AECRSD_2022}, ISLES 2015 \cite{maier2017isles}) having different modality sets, pathologies, and hospitals, summarized in Table~\ref{tab:datasetTable}. All datasets were skull stripped, resampled to 1x1x1 $mm^{3}$, and z-score normalized per modality. BRATS labels were merged into a single class for binary segmentation, aligning with other datasets.

\begin{table*}[!ht]
\centering
\caption{Dataset details including modalities, pathologies, and number of patients.}
\label{tab:datasetTable}
\tiny
\resizebox{\textwidth}{!}{%

\begin{tabular}{|c|c|c|c|c|c|c|c|c|c|}
\hline
\rowcolor{gray!10}  
\bf Datasets & \bf PD & \bf FLAIR  &\bf  T1 &\bf T1c  &\bf T2 &\bf DWI &\bf Pathology &\begin{tabular}[c]{@{}c@{}} \#\bf Train \bf patient \end{tabular} & \begin{tabular}[c]{@{}c@{}}\#\bf Test \bf patient \end{tabular}  \\\hline

BRATS-Decathlon~\cite{bakas2017advancing} & \ding{55} & \checkmark   & \checkmark & \checkmark & \checkmark &\ding{55} &Tumor &444&40\\\hline

ATLAS V2.0 \cite{liew2022large}& \ding{55} & \ding{55} & \checkmark & \ding{55} &\ding{55} & \ding{55} &Stroke lesion & 459 &196\\\hline

MSSEG \cite{commowick2018objective}&  \checkmark&  \checkmark& \checkmark &\checkmark  &  \checkmark&\ding{55}  &\begin{tabular}[c]{@{}c@{}} Sclerosis lesions \end{tabular} &37 &16 \\\hline

ISLES 2015 \cite{maier2017isles}& \ding{55} &\checkmark  &\checkmark  & \ding{55} &\checkmark  & \checkmark&Stroke lesion &20 & 8\\\hline

WMH \cite{AECRSD_2022}& \ding{55} & \checkmark & \checkmark & \ding{55} &\ding{55}  & \ding{55} &\begin{tabular}[c]{@{}c@{}}White matter\\ hyperintensity \end{tabular} &42 &18\\\hline

\end{tabular}
}
\end{table*}

Our method was implemented using Python 3.10.12 and PyTorch 2.1.0 and trained on an NVIDIA A40 GPU (48 GB VRAM). Each session ran for 400 epochs with a batch size of 4 and an input patch size of 128x128x128. We used the Adam optimizer~\cite{kingma2014adam} (learning rate 0.001, \(\beta_{1} = 0.9\), and \(\beta_{2} = 0.999\)) without schedulers. Simple random 90-degree rotations were applied for data augmentation, and training across all five datasets took about 20 hours with NVIDIA's Automatic Mixed Precision (AMP). During the inference phase, we employed MONAI’s~\cite{cardoso2022monai} sliding window technique. 

We compare the proposed strategy against popular CL strategies, including EWC~\cite{kirkpatrick2017overcoming}, SI~\cite{zenke2017continual}, LwF~\cite{radio3}, MiB~\cite{cermelli2020modeling}, TED~\cite{zhu2024boosting}, Replay~\cite{rolnick2019experience}, and GDumb~\cite{prabhu2020gdumb} in 3D-UNet. 
To establish baselines, we report lower bound performance with naive, and upper bounds with cumulative and joint training. Naive corresponds to traditional fine-tuning on new datasets, joint training uses all datasets simultaneously, and cumulative training sequentially incorporates all previous data. 
Experiments were conducted using Avalanche 0.6.0 framework~\cite{avalanche}. The buffer size was set to $200$ for Replay and GDumb. Regularization factors ($\alpha$ in LwF and $\lambda$ in SI and EWC) were tuned within {0.5, 1.0, 1.5, 2.0}, and $\tau$ was fixed at $2$. Hyper-parameters in MiB and TED were followed from respective papers. For our method, we set $\beta=0.8$, $\alpha_{max}=0.6$, and $e=4$. 
For domain-conditioned MoE, we consider maximum modalities as $m=6$ (PD, FLAIR, T1, T1c, T2, DWI) and pathology as $d=4$ (Tumor, Stroke lesion, Sclerosis lesions, White matter hyperintensity). An example of the binary domain-conditioned token ($\textbf{I}^{d+m}$) for a sample with FLAIR and T1 modality and ``stroke lesion" pathology would be [0, 1, 1, 0, 0, 0, 0, 1, 0, 0].
We tested on four dataset sequences: S1 (high to low dataset size: \{BRATS, ATLAS, MSSEG, ISLES, WMH\}), S2 (descending modality count: \{MSSEG, BRATS, ISLES, WMH, ATLAS\}), S3 (low to high dataset size: \{ISLES, WMH, MSSEG, BRATS, ATLAS\}), and S4 (ascending modality count: \{ATLAS, WMH, ISLES, BRATS, MSSEG\}).

In CL with $P$ datasets, training occurs sequentially across $P$ sessions, with the model evaluation after each session on the test sets of $P$ datasets. Consequently, for $P$ datasets, this process generates a $(P\times P)$ train-test matrix, where each cell $p_{ij}$ indicates the DSC on the test set of $D_j$ after sequential training from $D_1$ to $D_i$. 
CL metrics, including backward transfer (BWT) \cite{diaz2018don} and forward transfer (FWT) \cite{ozgun2020importance}, are derived from this matrix. We also report average performance (AVG), which measures DSC across all datasets after the $P^{th}$ session \cite{lopez2017gradient}, and the Incremental Learning Metric (ILM), i.e. the average of cells in the lower triangle of the matrix including diagonals, reflecting incremental learning capability \cite{diaz2018don}. Higher values of these metrics indicate better performance.


\begin{table*}[!t]
\centering
\caption{Method comparison (\textcolor{red}{best}, \textcolor{blue}{2nd best} in $\mathcal{B}$-free). $\mathcal{B}$: buffer; Cum.: Cumulative.}
\label{tab:resultsTableAll}
\tiny
\resizebox{\textwidth}{!}{%
\begin{tabular}{|c|l|cccc|cccc|cccc|cccc|}
\hline
\rowcolor{gray!10}
& 
& \multicolumn{4}{c|}{\textbf{S1}} 
& \multicolumn{4}{c|}{\textbf{S2}} 
& \multicolumn{4}{c|}{\textbf{S3}} 
& \multicolumn{4}{c|}{\textbf{S4}} 
\\ \cline{3-18}

\rowcolor{gray!10}
\multirow{-2}{*}{$\mathcal{B}$} 
& \multirow{-2}{*}{\textbf{Method}} 
& \textbf{AVG} & \textbf{ILM}& \textbf{BWT} & 
\textbf{FWT} 
& \textbf{AVG} & \textbf{ILM}& \textbf{BWT} & 
\textbf{FWT} 
& \textbf{AVG} & \textbf{ILM}& \textbf{BWT} & 
\textbf{FWT} 
& \textbf{AVG} & \textbf{ILM}& \textbf{BWT} & 
\textbf{FWT} 
\\ \hline

\multirow{4}{*}{\ding{51}} 
& Joint 
& 67.62 & - & - & - 
& 67.96 & - & - & - 
& 66.93 & - & - & - 
& 64.36 & - & - & - 
\\

& Cum.  
& 62.37 & 67.40 & -1.60 & 29.83 
& 69.20 & 73.04 & 0.05 & 21.46 
& 67.12 & 67.54 & -0.09 & 24.31 
& 66.79 & 61.04 & -1.33 & 16.31 
\\

& GDumb 
& 50.43 & 58.74 & -1.33 & 28.27 
& 50.63 & 62.42 & -11.89 & 21.25 
& 56.63 & 63.00 & -1.59 & 22.35 
& 58.10 & 52.69 & -5.42 & 17.14 
\\

& Replay 
& 67.09 & 68.62 & -4.98 & 33.08 
& 70.83 & 74.36 & -0.33 & 22.59 
& 70.68 & 66.79 & 3.93 & 19.71 
& 65.68 & 63.03 & 1.11 & 8.15 
\\ \hline

\multirow{7}{*}{\ding{55}} 
& Naive 
& 15.73 & 33.64 & -54.14 & 22.37 
& 23.43 & 37.36 & -54.16 & 17.94 
& 24.78 & 39.49 & -41.01 & 12.06 
& 36.66 & 39.58 & -38.69 & 14.73 
\\

& MiB 
& 26.89 & 41.80 & -45.06 & 23.52 
& 24.39 & 38.35 & -53.03 & \textcolor{red}{23.22} 
& 31.28 & 39.37 & -39.83 & 12.15 
& 39.47 & 39.43 & -35.96 & \textcolor{red}{18.57} 
\\

& TED 
& 31.49 & 44.08 & -40.86 & 26.92 
& 25.76 & 37.63 & \textcolor{blue}{-52.26} & 17.46 
& 33.44 & 38.46 & \textcolor{blue}{ -28.91} & 9.84 
& 42.08 & 43.69 & -32.92 & \textcolor{blue}{16.20 }
\\

& LwF 
& 29.97 & 41.18 & -45.15 & 22.37 
& 18.16 & 36.05 & -57.54 & 21.18 
& 24.82 & 38.24 & -40.10 & 14.31 
& 39.71 & 39.68 & -38.10 & 15.70 
\\

& SI 
& \textcolor{blue}{43.27} & \textcolor{blue}{51.69} & \textcolor{blue}{-25.07} & \textcolor{red}{31.31} 
& 13.32 & 36.83 & -52.69 & 18.00 
& 31.57 & 41.16 & -37.79 & \textcolor{blue}{15.56 }
& \textcolor{blue}{47.06} & \textcolor{blue}{43.87} & \textcolor{blue}{-25.94} & 13.68 
\\

& EWC 
& 26.48 & 39.04 & -45.30 & 21.05 
& \textcolor{blue}{26.78} & \textcolor{blue}{39.84} & -52.89 & 20.15 
& \textcolor{blue}{34.19} & \textcolor{blue}{41.55} & -37.65 & 13.97 
& 39.93 & 40.06 & -34.97 & 14.37 
\\

& Ours 
& \textcolor{red}{54.31} & \textcolor{red}{56.46} & \textcolor{red}{-16.46} & \textcolor{blue}{30.73} 
& \textcolor{red}{32.93} & \textcolor{red}{51.11} & \textcolor{red}{-27.28} & \textcolor{blue}{22.71} 
& \textcolor{red}{35.85} & \textcolor{red}{46.13} & \textcolor{red}{-21.09} & \textcolor{red}{21.53} 
& \textcolor{red}{50.67} & \textcolor{red}{48.54} & \textcolor{red}{-21.37} & 16.00 
\\ \hline

\end{tabular}
}
\end{table*}

\subsection{Results and discussion}\label{sec:resultsSec}

\textbf{Performance comparison with others:} For the considered medical applications, the primary concern will not be on improving zero-shot performance (FWT) but rather on minimizing forgetting (BWT) and enhancing the average DSC of the model (AVG and ILM). While FWT is reported for completeness, our analysis emphasizes AVG, ILM, and BWT.
Table~\ref{tab:resultsTableAll} presents the AVG, ILM, BWT, and FWT values for all methods across sequences S1, S2, S3, and S4. Across all sequences, CL approaches (GDumb, Replay, MiB, TED, LwF, SI, EWC, and the proposed method) mostly outperform naive training, highlighting the importance of mechanisms to mitigate CF in UNet-based segmentation tasks. Further, as expected, approaches storing past data partially (Replay, GDumb) or fully (cumulative, joint training) show higher performance compared to methods (naive, MiB, TED, LwF, SI, EWC, and the proposed approach) with no access to past exemplars.
When comparing the proposed method to other buffer-free approaches (MiB, TED, LwF, SI, EWC), it consistently achieves superior performance in all the sequences S1, S2, S3, and S4. Unlike these existing CL methods, which penalize large deviations from previously learned weights through response-level regularization terms in the training loss, the proposed approach introduces a drift-based dynamic penalization factor along with a latent-level regularization. This drift-based dual distillation allows for more effective mitigation of CF. 
The proposed method shows a positive gain in (AVG, ILM, and BWT) over the best performance achieved among state-of-the-art buffer-free approaches (blue colored in Table~{\ref{tab:resultsTableAll}}). Specifically, we observe an improvements of (25.51\%, 9.23\%, 34.34\%) in S1, (22.96\%, 28.29\%, 47.80\%) in S2, (4.85\%, 11.02\%, 27.05\%) in S3, and (7.67\%, 10.65\%, 17.62\%) in S4. 

\begin{figure*}[t]
\centering
\includegraphics[scale=0.2]{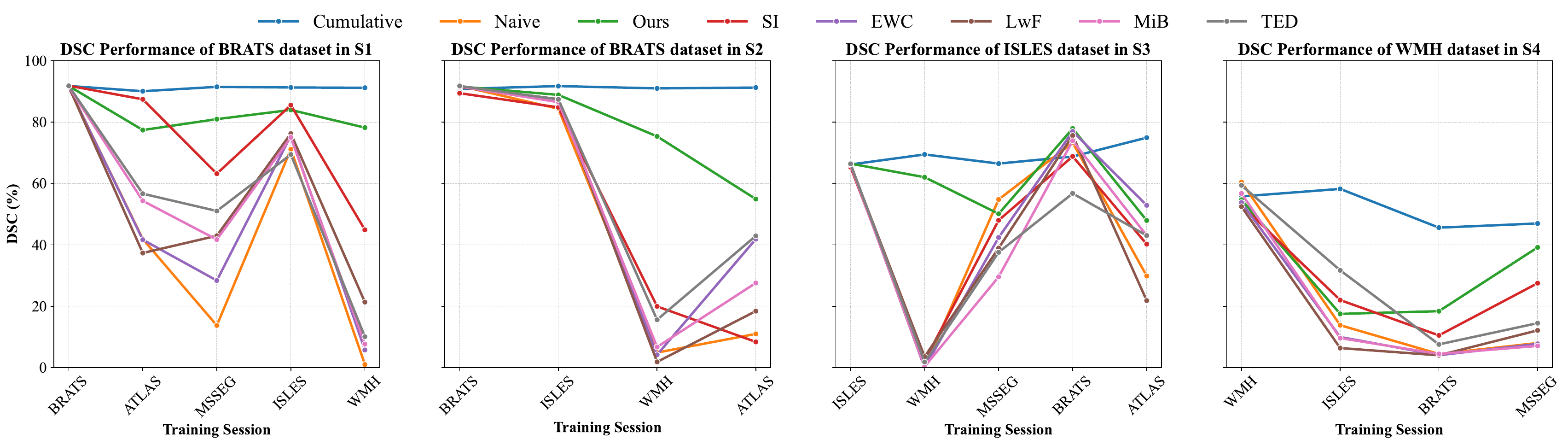}  
\caption{DSC on already learned datasets upon learning new datasets in S1-S4}
\label{fig:firstDatasetoverTime}
\end{figure*}

\textbf{Performance of a dataset in different sessions:} We closely analyze the CL model's performances on the first/second dataset upon learning other datasets in a given sequence. Specifically, Fig.~{\ref{fig:firstDatasetoverTime}} shows DSC for BRATS in S1 and S2, ISLES in S3, and WMH in S4, with cumulative training included for reference. While cumulative training offers stable results, it requires access to all previous datasets, which is impractical in real-world scenarios. The naive approach shows significant DSC degradation, with high standard deviations of 33.91 (S1) and 40.23 (S2), 26.86 (S3), and 22.64 (S4) reflecting instability. In contrast, our CL strategy maintains stability, with much lower standard deviations of 5.19 (S1), 14.51 (S2), 11.01 (S3), and 15.50 (S4) highlighting its increased robustness against forgetting. Other buffer-free CL methods (TED, MiB, SI, EWC, LwF) show better performance than naive training (Table~{\ref{tab:resultsTableAll}}) but still exhibit instability in DSC, with standard deviations of (26.83, 28.80, 17.86, 31.14, 25.98) for S1, (31.72, 36.75, 36.75, 35.85, 40.10) for S2, (22.12, 26.41, 23.84, 26.60, 26.85) for S3, and (20.00, 21.61, 15.51, 20.26, 19.69) for S4. While these methods perform well for natural images, their effectiveness is limited in brain MRI segmentation under domain shifts. In contrast, our approach delivers better stability and mitigates CF effectively. 

\begin{figure}[t]
    \centerline{\includegraphics[width=\columnwidth]{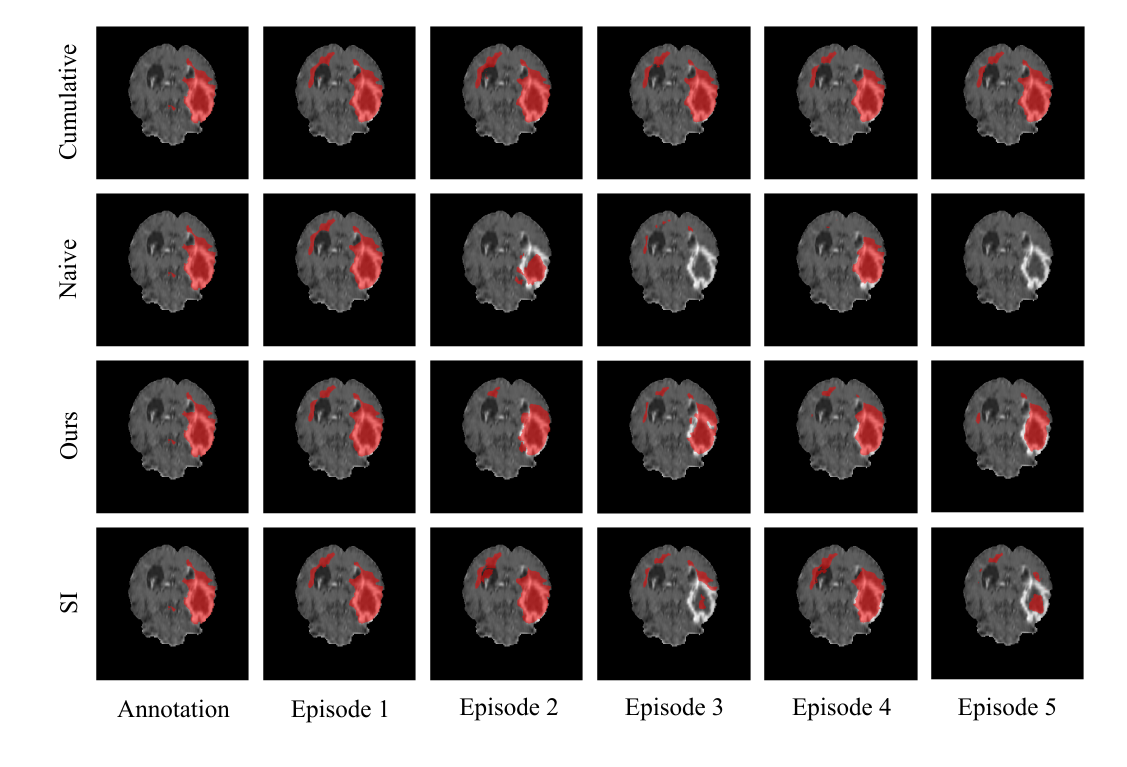}}
\caption{Qualitative comparison for a patient in BRATS over subsequent episodes in S1.}
\label{fig:results-S1}
\end{figure}

\textbf{Impact of dataset orders:} We study the impact of different sequences on overall performance. We analyze AVG, ILM, and BWT by best performing other CL methods and proposed approach for S1-S4 in Table~{\ref{tab:resultsTableAll}}.  
We can see that the best AVG is poor in S2 (32.93) and S3 (35.85) as compared to that in S1 (54.31) and S4 (50.67). Notably, all methods (except EWC) showed performance degradation when ATLAS (a single-modality dataset) was introduced later in sequences (S2, S3), adversely impacting the generalization of previously acquired knowledge in the model. This occurs because ``modality dropping", a critical generalization technique, cannot be applied to ATLAS due to its single-modality nature. Consequently, learning ATLAS in the later stages negatively impacts the model’s prior generalization capability. In contrast, when we learn datasets with fewer modalities at the start of sequences (S1 and S4), their negative effect is covered at later stages when we learn datasets with more modalities. 

\textbf{Qualitative Results:}
Fig. \ref{fig:results-S1} shows how tumor segmentation of a sample learned in the first episode evolves over subsequent episodes across different approaches. We compare segmentation results on a randomly selected slice from a BRATS patient using cumulative, naive, our proposed method, and the best-performing buffer-free strategy (SI) in S1. The cumulative method maintains consistent tumor predictions but introduces substantial false positives, particularly in the upper-left brain region, indicating limited generalization. The naive approach, which lacks CL mechanism, initially performs well but degrades rapidly across episodes, ultimately losing nearly all segmentation capability. While the SI method performs reasonably early on, it suffers a significant decline over time, ultimately failing to segment much of the tumor in the final episode, indicating poor knowledge retention. In contrast, our approach initially produces some false positives but progressively refines over time. By the final episode, it preserves accurate tumor regions while minimizing misclassifications, showing strong retention and adaptability. These results highlight the method's effectiveness in mitigating CF in continual segmentation tasks.

\begin{table*}[t]
\centering

\caption{Ablation study. \textbf{Bold} refers to the proposed approach. }
 \label{tab:abalationTable}
 \tiny
\resizebox{\textwidth}{!}{%
\begin{tabular}{|c|ccc|cccc|cccc|}
\hline
\rowcolor{gray!10} 
&\multicolumn{3}{c|}{} & \multicolumn{4}{c|}{\textbf{S1}} & \multicolumn{4}{c|}{\textbf{S2}}  \\ \hline

\rowcolor{gray!10} 
\begin{tabular}[c]{@{}c@{}}\textbf{Method} \end{tabular} 
& $\mathcal{L}^{\text{KLD}}$ & $\mathcal{L}^{\text{Cosine}}$ & \textbf{MoE}
& \textbf{ACC}$\uparrow$ & \textbf{ILM}$\uparrow$ & \textbf{BWT}$\uparrow$ & \textbf{FWT}$\uparrow$  
& \textbf{ACC}$\uparrow$ & \textbf{ILM}$\uparrow$ & \textbf{BWT}$\uparrow$ & \textbf{FWT} $\uparrow$
 \\\hline

\begin{tabular}[c]{@{}c@{}} Naive \end{tabular} 
&\multicolumn{1}{c}{-}&\multicolumn{1}{c}{-}&\multicolumn{1}{c|}{-}
&\multicolumn{1}{c}{15.73}&\multicolumn{1}{c}{33.64}&\multicolumn{1}{c}{-54.14}&\multicolumn{1}{c|}{22.37}
&\multicolumn{1}{c}{23.43}&\multicolumn{1}{c}{37.36}&\multicolumn{1}{c}{-54.16}&\multicolumn{1}{c|}{17.94}
 \\\hline

&\multicolumn{1}{c}{\checkmark}&\multicolumn{1}{c}{\ding{55} }&\multicolumn{1}{c|}{\ding{55}}
&\multicolumn{1}{c}{41.29}&\multicolumn{1}{c}{47.16}&\multicolumn{1}{c}{-11.31}&\multicolumn{1}{c|}{24.37}&\multicolumn{1}{c}{13.79}&\multicolumn{1}{c}{38.81}&\multicolumn{1}{c}{-30.43}&\multicolumn{1}{c|}{23.13}\\

&\multicolumn{1}{c}{\ding{55}}&\multicolumn{1}{c}{\checkmark}&\multicolumn{1}{c|}{\ding{55}}
&\multicolumn{1}{c}{24.31}&\multicolumn{1}{c}{37.85}&\multicolumn{1}{c}{-51.90}&\multicolumn{1}{c|}{25.66}&\multicolumn{1}{c}{33.23}&\multicolumn{1}{c}{42.25}&\multicolumn{1}{c}{-48.39}&\multicolumn{1}{c|}{25.44}\\

&\multicolumn{1}{c}{\checkmark}&\multicolumn{1}{c}{\checkmark}&\multicolumn{1}{c|}{\ding{55}}
&\multicolumn{1}{c}{49.20}&\multicolumn{1}{c}{52.03}&\multicolumn{1}{c}{-18.82}&\multicolumn{1}{c|}{32.34}&\multicolumn{1}{c}{29.62}&\multicolumn{1}{c}{45.58}&\multicolumn{1}{c}{-15.89}&\multicolumn{1}{c|}{22.38}\\

&\multicolumn{1}{c}{\ding{55}}&\multicolumn{1}{c}{\ding{55}}&\multicolumn{1}{c|}{$\textbf{I}^{d}$}
&20.21&35.32&-56.74&22.22
&25.05&37.88&-56.01&23.17
\\

Our&\multicolumn{1}{c}{\ding{55}}&\multicolumn{1}{c}{\ding{55}}&\multicolumn{1}{c|}{$\textbf{I}^{m}$}
&22.19&38.59&-51.06&25.23
&21.50&38.13&-55.61&21.27
\\

&\multicolumn{1}{c}{\ding{55}}&\multicolumn{1}{c}{\ding{55}}&\multicolumn{1}{c|}{$\textbf{I}^{d+m}$}
&\multicolumn{1}{c}{22.91}&\multicolumn{1}{c}{42.36}&\multicolumn{1}{c}{-45.83}&\multicolumn{1}{c|}{25.65}&\multicolumn{1}{c}{23.13}&\multicolumn{1}{c}{41.40}&\multicolumn{1}{c}{-51.21}&\multicolumn{1}{c|}{26.09}\\

&\multicolumn{1}{c}{\checkmark}&\multicolumn{1}{c}{\checkmark}&\multicolumn{1}{c|}{$\textbf{I}^{d}$}
&47.29&54.94&-20.59&29.63
&26.76&48.92&-35.68&22.14
\\

&\multicolumn{1}{c}{\checkmark}&\multicolumn{1}{c}{\checkmark}&\multicolumn{1}{c|}{$\textbf{I}^{m}$}
&40.28&48.55&-33.07&21.10
&13.17&44.32&-40.23&24.71
\\

&\multicolumn{1}{c}{\checkmark}&\multicolumn{1}{c}{\checkmark}&\multicolumn{1}{c|}{$\textbf{I}^{d+m}$}
&\multicolumn{1}{c}{\bf 54.31}&\multicolumn{1}{c}{\bf 56.46}&\multicolumn{1}{c}{\bf -16.46}&\multicolumn{1}{c|}{ \bf 30.73}&
\textbf{32.93} & \textbf{51.11} & \textbf{-27.28} & \textbf{22.71} 
\\\cline{1-12}

\multicolumn{4}{|c|}{Without modality drop strategy}
&\multicolumn{1}{c}{34.80}&\multicolumn{1}{c}{37.65}&\multicolumn{1}{c}{-50.15}&\multicolumn{1}{c|}{13.41}&\multicolumn{1}{c}{18.25}&\multicolumn{1}{c}{35.08}&\multicolumn{1}{c}{-58.58}&\multicolumn{1}{c|}{15.92}
\\\hline

\end{tabular}%
 }
\end{table*}

\subsection{Ablation study}
We present the results of our ablation study in Table~\ref{tab:abalationTable}, showing that all modules of our approach, both individually and in combination, outperform naive UNet training for variable-modality brain MRI segmentation. Notably, our dynamic regularization coefficient for response-based regularization ($\mathcal{L}^{\text{KLD}}$) outperforms the static coefficient in LwF (Table~\ref{tab:resultsTableAll}). While LwF achieves ILM scores of 41.18 and 36.05 in S1 and S2, respectively, our dynamic setting results in higher ILM scores of 47.16 and 38.81. This approach not only improves performance but also eliminates the need to manually select suitable $\alpha$ for a given dataset sequence. We also observe that removing the random dropping of modalities harms performance, as the model learns undesired association of a dataset with a fixed modality set, preventing it from developing generalized representations, which are crucial for learning diverse datasets within a single model.
Next, we analyze the contribution of each module in our approach. In S1, response-based regularization outperforms latent-based regularization, while the reverse is true in S2. However, their combination, the dual-distillation consistently outperforms both in S1 and S2, supporting its inclusion in the proposed strategy. Additionally, the domain-conditioned MoE ($\textbf{I}^{d+m}$) outperforms naive UNet in both experiments. Studying the joint token ($\textbf{I}^{d+m}$) against the individual modality-only ($\textbf{I}^{m}$) and disease-only ($\textbf{I}^{d}$) tokens in presence or absence of dual distillation, we find that their combination outperforms the individual tokens. Further, when the joint token is combined with dual-distillation, average DSC improves significantly: ILM increases from 52.03 to 56.46 in S1 and from 45.58 to 51.11 in S2, reflecting gains of 8.51\% and 12.13\%, respectively. These results demonstrate the complementary benefits of dual-distillation and domain-conditioned MoE, highlighting the effectiveness of our hybrid CL strategy.

\subsection{Limitations and future work}

This work takes an initial step toward buffer-free CL for brain lesion segmentation across variable modalities with a simple training setup. 
While it outperforms naive training and existing CL strategies, it still underperformed compared to models trained separately for each dataset, reflecting the trade-offs of a unified CL approach. 
Enhanced training setups, such as extended schedules or hyperparameter tuning, could improve results. 
Although buffer-free learning is practical, it may impair long-term retention and exacerbate forgetting in highly diverse datasets.
Future work will explore privacy-preserving latent replay, adaptive learning schedules, optimized architectures, and scaling to more diverse datasets.

\section{Conclusion}\label{sec:conc}

This work tackles brain lesion segmentation in multi-modal MRI under real-world constraints, where datasets arrive sequentially with varying modalities. We introduced a hybrid CL strategy that enables a single segmentation model to learn from diverse datasets without requiring all data at once. Using two-stage distillation and a mixture-of-experts mechanism, our approach reduces catastrophic forgetting while preserving performance on previously seen data. Experiments on five diverse brain MRI datasets in four different sequences validate its effectiveness in handling variations across modalities, hospitals, and pathologies. This study marks a step toward practical, scalable multi-modal MRI segmentation, demonstrating the potential of CL in neuroimage analysis.

\begin{credits}
\subsubsection{\ackname} The authors gratefully acknowledge the computational and data resources provided by  \href{https://www.lrz.de}{the Leibniz Supercomputing Centre}. Also, the authors gratefully acknowledge the scientific support and HPC resources provided by the Erlangen National High-Performance Computing Center (NHR@FAU) of the Friedrich-Alexander-Universität Erlangen-Nürnberg (FAU) under the NHR project “b213da.” NHR funding is provided by federal and Bavarian state authorities. NHR@FAU hardware is partially funded by the German Research Foundation (DFG) – 440719683. Also, this work was supported by the German Research Foundation (Deutsche Forschungsgemeinschaft, DFG) under the grant no. 417063796.

\subsubsection{\discintname}
 The authors have no competing interests to declare that are
relevant to the content of this article. 
\end{credits}

\bibliographystyle{splncs04}
\bibliography{2}

\end{document}